\documentclass[10pt,aps.prl,twocolumn,superscriptaddress,floatfix]{revtex4-1} 

\usepackage{amsmath}

\usepackage{graphics}
\usepackage{color}
\usepackage{caption}
\usepackage{subcaption}
\usepackage{float}
\usepackage{array}
\usepackage{ulem}
\usepackage{amssymb}
\usepackage{bm}

\begin{document}

\title{Dy$^{3+}$-doped Yttrium Complex Molecular Crystals for Two-color Thermometry in Heterogeneous Materials}
\author{Benjamin R Anderson}
\address{Applied Sciences Laboratory, Institute for Shock Physics, Washington State University,
Spokane, WA 99210-1495}
\email{branderson@wsu.edu}
\author{Ray Gunawidjaja}
\address{Applied Sciences Laboratory, Institute for Shock Physics, Washington State University,
Spokane, WA 99210-1495}
\author{Hergen Eilers}
\address{Applied Sciences Laboratory, Institute for Shock Physics, Washington State University,
Spokane, WA 99210-1495}
\email{eilers@wsu.edu}
\date{\today}

\begin{abstract}
We develop Dy$^{3+}$-doped yttrium complexes for use as two-color thermometry (TCT) phosphor molecular crystals in heterogeneous materials. These complexes include: Dy:Y(acac)$_3$(phen), Dy:Y(hfa)$_3$(DPEPO), Dy:Y(4-BBA)$_3$(TPPO), Dy:Y(acac)$_3$, and Dy:Y(acac)$_3$(DPEPO), where the Dy/Y ratio is 1:9. We characterize the materials' photoluminescence at different temperatures to determine the TCT calibration parameters and the degree to which thermal quenching influences the emission. From this data we observe a link between the excited state lifetime at room temperature and the degree to which the material is susceptible to thermal quenching (i.e. materials having long room temperature lifetimes are more resistant to thermal quenching than materials with short room temperature lifetimes). Of the five complexes tested we find that Dy:Y(acac)$_3$(DPEPO) has the best thermal performance, with the most likely source of improvement being DPEPO's compact rigid structure. This rigidity helps with energy transfer to the Dy$^{3+}$ ion, suppresses non-radiative loss modes, and reduces exciplex formation.
\end{abstract}

\maketitle

\section{Introduction}
Over the past several decades much work has been dedicated to elucidating the mechanisms of mechanical shock induced ignition of plastic bonded explosives (PBXs)\cite{Chen14.01,Chen14.02,Pokharel14.01,An13.01,Wu11.01,Rae02.01,Gupta95.01,Mellor93.01,Field92.01,Field92.02,Field82.01,Winter75.01,Chaudhri74.01}. PBXs are heterogeneous materials consisting of energetic molecular crystals embedded in a polymer matrix, which sometimes also contain taggants/markers, friction-generating grit, antioxidants, and plasticizers.  Through this research it has been determined that during shock compression, hot spots form which, under the right conditions can lead to the ignition of the PBX.

These hot spots form through several different mechanisms, including: micro-fractures, plastic deformation \cite{Pokharel14.01,Winter75.01,Field82.01}, pore collapse \cite{Chaudhri74.01}, binder/crystal interface friction, and crystal/crystal interface friction. However, while these mechanisms have been determined, it is currently unknown which mechanisms are most important given a specific material composition and specific shock loading conditions \cite{Field82.01}. In order to answer this question we are developing an imaging thermometry technique -- based on lanthanide luminescence -- for potential use in combination with microstructural imaging of heterogeneous materials during shock compression. By performing simultaneous structural and thermal imaging of heterogenous materials we hope to be able to determine how different loading conditions lead to hot spot formation.

While the long term goal is to combine both thermal and structural imaging in shock experiments, the first step is to develop heterogeneous samples suitable for thermal imaging. To achieve this we are developing lanthanide based molecular crystals and dyes which can be dispersed/doped into a polymer to provide a photoluminescent heterogeneous material. Thermometric imaging is then achieved through measuring the temperature-dependent photoluminesence properties of the lanthanides.

Thermometry using lanthanide photoluminescence is a broad field with a wide range of experimental techniques and host materials (for an overview of this field see Ref. \cite{Kontis07.01,Khalid08.01,Allison97.01}). For our specific application we have chosen the technique of two-color ratio thermometry \cite{Juboori13.01,Heyes06.01,Kontis07.01,Kontis02.01,Heyes06.01}  as it is simple to implement in imaging experiments and provides good temporal resolution. This technique involves measuring the temperature dependent intensity ratio of two emission peaks, from which the temperature can be determined. Specifically, we have chosen to use Dy$^{3+}$-doped materials as Dy$^{3+}$'s energy levels have two advantageous features for two-color thermometry (TCT). First, Dy$^{3+}$ has two closely spaced energy levels (${}^4F_{9/2}$ and ${}^4I_{15/2}$), which are separated by an energy splitting of $\Delta E\approx$ 0.115 eV (928 cm$^{-1}$) and second, the splitting between the highest Stark level of the ${}^6H_{15/2}$ ground state and the ${}^4F_{9/2}$ excited state is $\approx 0.97$ eV (7850 cm$^{-1}$)(which means relaxation will be predominately radiative) \cite{Bunzli10.01}. 

When the Dy ion is excited by an appropriate wavelength of light it is excited into the ${}^4F_{9/2}$ energy level (population $n_1$) from which it can be thermally excited into the ${}^4I_{15/2}$ energy level (population $n_2$). The ratio of the two populations is given by the Boltzmann distribution as

\begin{align}
\frac{n_2}{n_1}=e^{-\Delta E/kT}.
\end{align}
This ratio of populations can be optically determined, as the intensity ratio of the ${}^4F_{9/2}\rightarrow {}^6H_{15/2}$ and ${}^4I_{15/2}\rightarrow {}^6H_{15/2}$ optical transitions are proportional to the population ratio
\begin{align}
\frac{I_2}{I_1}&\propto\frac{n_2}{n_1},
\\ &=Ae^{-\Delta E/kT}, \label{eqn:fitfunc}
\end{align}
where $A$ is a factor relating the populations to the measured intensity spectra. Therefore by determining both $A$ and $\Delta E$  in Equation \ref{eqn:fitfunc} we can use a measured intensity ratio to determine the temperature of the Dy ions.

\begin{figure}
\centering
\includegraphics{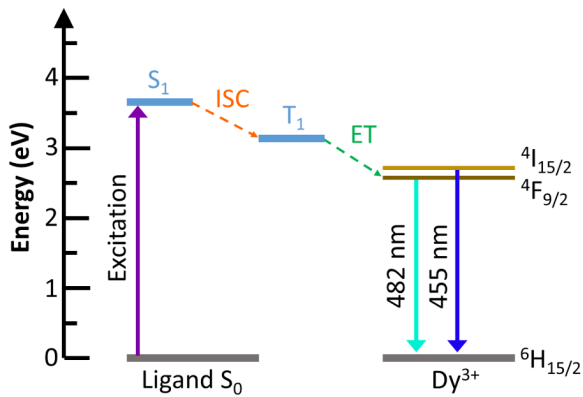}
\caption{Schematic of energy transfer between the ligands and the Dy ion. The ligand first absorbs the laser light (355 nm in our case) and is excited into the first excited singlet state, $S_1$. The ligand then undergoes inter-system crossing into the triplet state, $T_1$, from which the energy is transferred to the Dy ion either through dipole-dipole interactions or electron exchange. Note that in general the excitation wavelength depends on the ligand's $S_1$ state and therefore we leave the excitation wavelength ambiguous.}
\label{fig:ET}
\end{figure}

While Dy's energy level spacing makes it a good lanthanide for TCT, its performance is limited by its intrinsically low molar absorbance \cite{Bunzli10.01}, which means that its fluorescence conversion efficiency is low. One method of overcoming this deficiency is to attach ligands (which have large absorption coefficients) to the Dy ions, which can then absorb incident light and transfer the energy to the Dy ion \cite{Sager65.01,Dexter53.01,Xin04.01,Latva97.01,Bunzli10.01,Bunzli89.01,Eliseeva10.01}, essentially using these ligands as ``antennas'' for the Dy ion. Figure \ref{fig:ET} demonstrates a schematic of this energy transfer, with the process as follows: (1) the ligand absorbs the laser light and is excited into the $S_1$ state, (2) the ligand relaxes into the lowest triplet state ($T_1$) via inter-system crossing, (3) the energy is transferred to the lanthanide's primary emissive state via: dipole-dipole interactions \cite{Forster65.01,Forster59.01,Dexter53.01}, dipole-quadrapole interactions \cite{Dexter53.01}, excitons \cite{Dexter51.01,Heller51.01} or electron exchange \cite{Dexter53.01,Adronov00.01}, and (4) the excited lanthanide ion relaxes emitting a photon.

Currently there is no universal law for determining which ligand-lanthanide combinations provide the best energy transfer to the lanthanide ion. However, Latva \textit{et al.} observed -- in a study of many different ligand-lanthanide combinations -- that efficient energy transfer occurs when the energy difference between the ligand's lowest triplet state and the lanthanide's main fluorescent state is $>0.23$ eV \cite{Latva97.01}. Here we consider six different ligands with their reported energy levels tabulated in Table \ref{tab:ligand} along with the energy difference between their lowest triplet state and Dy$^{3+}$'s ${}^4F_{9/2}$ energy level. Note that the $S_1$ and $T_1$ energies of each ligand are found to vary in the literature \cite{Feng09.01,Sager65.01,Congiu13.01,Feng09.01,Xu06.01,Xin06.01,Peng05.01,Biju13.01,Zhang12.01,Xin04.01} with the values reported in Table \ref{tab:ligand} being average values of these reports.

\begin{table*}
 \caption{Approximate energies for the $S_1$ and $T_1$ energy levels for each ligand and the energy difference between the ligand's lowest triplet state and Dy$^{3+}$'s ${}^4F_{9/2}$ energy level.}
 \label{tab:ligand}
 \begin{tabular}{ccccc}
 \hline
  \textbf{Ligand}   &  \textbf{$S_1$ Energy (eV)}  &  \textbf{$T_1$ Energy (eV)}   & $\Delta E$ (eV) & \textbf{Ref.} \\ \hline
  acac   &3.60  &3.13  & 0.55 &  \cite{Feng09.01,Sager65.01}   \\
  hfa    &4.25  &2.77  & 0.18 &  \cite{Congiu13.01}  \\ 
  phen   &3.26  &2.85  & 0.27 &  \cite{Feng09.01,Xin06.01,Peng05.01}   \\
  DPEPO  &3.94  &2.99  & 0.41 &  \cite{Biju13.01,Xu06.01,Congiu13.01}  \\
  4-BBA  &4.14  &2.95  & 0.37 &  \cite{Zhang12.01}  \\ 
  TPPO   &4.51  &3.01  & 0.43 &  \cite{Feng09.01,Xu06.01,Zhang12.01,Xin04.01}  \\ \hline
 \end{tabular}
\end{table*}

Looking at Table \ref{tab:ligand} we find that all ligands chosen, except hfa, satisfy Latva's empirical rule with $\Delta E>0.23$ eV. The reason for including hfa in our study is that hfa is the H-F substituted version of acac. Previously it has been shown that in some cases H-F substitution in a Dy-complex can lead to improved luminescence performance due to H-F substitution eliminating high-energy vibrations from the ligand framework \cite{Glover07.01}. Therefore we are interested in determining if the H-F substitution in acac will lead to improved luminescent performance, despite hfa not satisfying Latva's rule.

While ligands (with advantageous energy level spacings) can act as antennas for lanthanide ions, they also provide two additional benefits for our application. These benefits are: (1) ligands can be used to form Dy-doped organic molecular crystals, which act as solid inclusions in heterogeneous materials and (2) the ligands can be used to form a polymer soluble dye, which allows for the Ln-doped complex to be dispersed throughout a polymer. In the future, these two materials (both the molecular crystals and dye) will be combined in HTPB polymer to form heterogeneous materials for dynamic shock compression experiments. Having the TCT phosphors in both the molecular crystals and dispersed in the polymer should allow for spatially resolved thermometric imaging of both the crystals and polymer.

In this paper we consider five different TCT phosphor molecular crystals (based on Dy$^{3+}$-doped yttrium complexes), including: Dy:Y(acac)$_3$(phen) \cite{Anderson17.01}, Dy:Y(hfa)$_3$(DPEPO), Dy:Y(4-BBA)$_3$(TPPO), Dy:Y(acac)$_3$, and Dy:Y(acac)$_3$ (DPEPO), where the Dy/Y ratio is 1:9. We compare the performance of the four new Dy$^{3+}$-doped yttrium complexes and Dy:Y(acac)$_3$(phen) to determine the best phosphor for our eventual application: thermometry of heterogeneous samples during dynamic shock compression. Given this application there are five main performance parameters we consider:  (1) room temperature luminescence lifetime, (2) functional maximum temperature $T_M$ (which we define as the temperature at which thermal quenching cuts the emission intensity to 1\% of the room temperature level), (3) the intensity proportionality constant ($A$ in Equation \ref{eqn:fitfunc}), (4) the temperature resolution $\Delta T$, and (5) the maximum temperature sensitivity $S_M$. The importance of each of these parameters is as follows:

\begin{enumerate}
 \item The room temperature lifetime is a good performance parameter for two reasons: first, the lifetime is proportional to the quantum efficiency of the phosphor \cite{Bunzli10.01}, with longer lifetimes having higher efficiencies. Secondly, the phosphor lifetime limits the time range over which TCT measurements can be performed during a shock experiment. Since shock experiments typically occur over microseconds, and we are using a laser with a repetition rate of 10 Hz, we will only be able to use a single excitation pulse to excite the phosphor. The phosphor then has to luminescence over the whole duration of the shock event in order to make multiple TCT measurements.  

\item The functional maximum temperature $T_M$, is the temperature at which the intensity of the $\lambda_1$ peak is quenched to 1\% the room temperature value. At this intensity level the spectra are too noisy for accurate temperature calculations using our current spectroscopy system. Note that this definition of the functional maximum temperature depends on a specific spectroscopy system and can vary depending on the experimental setup used to measure the material's photoluminescence.

\item The intensity proportionality constant $A$, is a measure of the relative intensity of the higher energy transition's peak as compared to the lower energy transition's peak. In practical terms this means that as $A$ increases the higher energy transition's peak is brighter at room temperature. This directly affects the signal-to-noise ratio, with larger $A$'s providing improved signal-to-noise ratios.

\item The temperature resolution -- defined as \cite{Anderson17.01}
\begin{align}
\Delta T(T)=\sigma_r(T)\left(\frac{\partial r(T)}{\partial T}\right)^{-1}, \label{eqn:Tres}
\end{align}
where $r(T)$ is the integrated intensity ratio at temperature $T$ and $\sigma_r(T)$ is the experimental uncertainty in the ratio -- is a direct measure of the precision of the TCT phosphor and optical hardware. It determines how well two closely spaced temperatures can be resolved using a given TCT phosphor and optical setup.

\item The temperature sensitivity $S$, is defined as \cite{Brites12.01}

\begin{align}
S=\frac{100}{r(T)}\frac{\partial r(T)}{\partial T}, \label{eqn:sens}
\end{align}
and is given in units of \% K$^{-1}$. It determines the fractional change in the resolution as a function of temperature, with larger sensitivities implying larger ratio changes for a given change in temperature.

\end{enumerate}

\section{Method}
\subsection{Materials}

For our trial Dy$^{3+}$-doped yttrium complex based molecular crystals we choose the following five complexes:  Dy:Y(acac)$_3$, Dy:Y(acac)$_3$(phen), Dy:Y(acac)$_3$(DPEPO), Dy:Y(hfa)$_3$(DPEPO), and Dy:Y(4-BBA)$_3$(TPPO). In this section we briefly describe the synthesis of each complex, with each one prepared with a Dy-doping of 10 mol\%.

\subsubsection{Dy:Y(acac)$_3$}
We begin synthesis of Dy:Y(acac)$_3$ by first preparing a 10 mL aqueous solution of Dy(NO$_3$)$_3$.5H$_2$O/Y(NO$_3$)$_3$.6H$_2$O (0.625 M), which is then mixed with a solution of acetylacetone, acac (7.5 M), in 2.5 mL methanol. Next, a 2.5 mL aqueous solution of NaOH (1.5 M) is added to cause precipitation of the product, with the suspension being aged for 2-3 h. Afterwards the product is washed with deionized water and dried in vacuum oven at 80 $^\circ$C for $\approx 12$ hours.

\subsubsection{Dy:Y(acac)$_3$(phen)}

Dy:Y(acac)$_3$(phen) is prepared as follows: first a solution of acetylacetone (0.06 M) and 1,10-phenanthroline (0.02 M) in a DMF/methanol mixture (1:1 vol/vol ratio) is prepared. Once prepared we add an aqueous solution of potassium tert-butoxide (KOtBu) (0.06 M) at an equal volume ratio. To this mixture we then add another equal volume aqueous solution of Dy(NO$_3$)$_3$.5H$_2$O and Y(NO$_3$)$_3$.6H$_2$O such that the total concentration is 0.02 M. The resulting solution is stirred for 4-6 hours during which white precipitates begin to form. The precipitates are then isolated using centrifugation (6000 rpm for 3 min) and vacuum dried at 80 $^\circ$C for $\approx 12$ hours.

\subsubsection{Dy:Y(acac)$_3$(DPEPO)}
We prepare Dy:Y(acac)$_3$(DPEPO) by first preparing Dy(acac)$_3$, as described above, which is then mixed with DPEPO at an equimolar ratio. The mixture is subsequently dissolved in a methanol/DMF mixture (0.06 M) and refluxed overnight. The supernatant is decanted to give the solid product, which is finally dried in a vacuum oven at 80 $^\circ$C for $\approx 12$ hours.

\subsubsection{Dy:Y(4-BBA)$_3$(TPPO)}
Dy:Y(4-BBA)$_3$(TPPO) is prepared using the method of Zhang \textit{et al.} \cite{Zhang12.01} in which a solution of Dy(NO$_3$)$_3$.5H$_2$O /Y(NO$_3$)$_3$.6H$_2$O (0.1 M) in 30 ml ethanol (0.1 M) is first prepared. To this solution we add 4-benzoylbenzoic acid (4-BBA)(1.8 M) and triphenylphosphine oxide (TPPO)(1.2 M) in 5 ml of DMF. To induce precipitation of the product we add a 30 ml aqueous solution of 0.8 M NaOH. The suspension is then boiled at 60-70 $^\circ$C for 2-3 h and the resulting product is washed with deionized water and dried in a vacuum oven at 80 $^\circ$C for $\approx 12$ hours.

\subsubsection{Dy:Y(hfa)$_3$(DPEPO)}
Dy:Y(hfa)$_3$(DPEPO) is prepared following the procedure of Chen \textit{et al.} \cite{Chen11.01}. We first dissolve Dy(NO$_3$)$_3$.5H$_2$O /Y(NO$_3$)$_3$.6H$_2$O (0.02 M), DPEPO (0.02 M), and hexafluoroacetylacetone, (hfa) (0.06 M) in a dimethylformamide /methanol mixture (1:1 vol/vol ratio). We then add an equivolume aqueous solution of KOtBu (0.06 M), which causes the product to precipitate. The suspension is allowed to age for 3 h, after which the precipitates are isolated using centrifugation (6000 rpm for 5 min). With the precipitates isolated we dry them in a vacuum over at 80 $^\circ$C for $\approx 12$ hours.

\subsection{Optical Measurements}
We perform temperature dependent emission spectroscopy of molecular crystals using a custom powder heater and fluorescence spectrometer \cite{Anderson17.01}. The spectrometer system consists of a frequency-tripled Nd:YAG laser (Continuum Powerlite, 355 nm, 8 ns, 10 Hz), various focusing and collection optics, and an Acton SpectraPro 2500i monochromator with attached PI-Max 3 ICCD camera. We also use a PMT attached to an Acton SpectraPro 2750i monochromator to perform lifetime measurements with the PMT output connected to a Tektronix DPO 4104 oscilloscope.

To heat the samples we use a custom built powder heater consisting of a solid cylinder of aluminum (2" D $\times$ 1" H) with a 1 cm diameter indentation placed in its center where the powder is placed. By surrounding the powder on the sides and bottom we are able to more evenly heat the powder. To heat the aluminum cylinder we use a 120 W canister heater, which is controlled by an Omega CN32PT-220 PID controller with a K-type thermocouple providing temperature feedback.


\section{Results}

\subsection{Room Temperature Luminescence}
\begin{figure}
 \centering
 \includegraphics{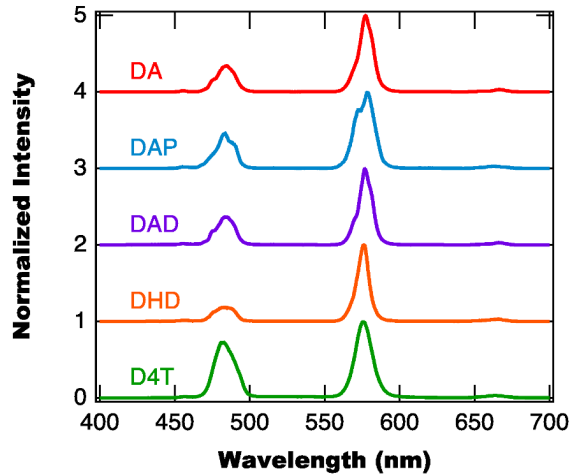}
 \caption{Normalized emission spectra of the five different materials at room temperature for an excitation wavelength of 355 nm. Table \ref{tab:peaks} tabulates the multipeak fit parameters for all five materials. DA = Dy:Y(acac)$_3$, DAP = Dy:Y(acac)$_3$(phen), DHD = Dy:Y(hfa)$_3$(DPEPO), D4T = Dy:Y(4-BBA)$_3$(TPPO), DAD = Dy:Y(acac)$_3$(DPEPO)}
 \label{fig:spec}
\end{figure}

We start our study of the spectral properties of the different Dy$^{3+}$-doped yttrium complexes by considering the room temperature emission spectra and the lifetimes of the ${}^4F_{9/2}$ and ${}^6I_{15/2}$ energy levels. Figure \ref{fig:spec} shows the normalized emission spectra of all five materials with each found to consist of broad emission peaks at $\approx455$ nm (${}^4I_{15/2}\rightarrow {}^6H_{15/2}$), $\approx 482$ nm (${}^4F_{9/2}\rightarrow {}^6H_{15/2}$), $\approx 575$ nm (${}^4F_{9/2}\rightarrow {}^6H_{11/2}$) and $\approx 660$ nm (${}^4F_{9/2}\rightarrow {}^6H_{9/2}$). While all five spectra in Figure \ref{fig:spec} are similar, there are differences in the peak structure with the precise peak locations depending on which ligands are attached.

To better understand the differences in the spectra shown in Figure \ref{fig:spec} we perform multi-Gaussian-peak fitting of each spectra and list the peak location and fractional peak area in Table \ref{tab:peaks}, where the fractional peak area is defined as

\begin{align}
\text{FA}_i\equiv \frac{A_i\sigma_i}{\sum\limits_i^N A_i\sigma_i},
\end{align}
where $A_i$ is the peak amplitude and $\sigma_i$ is the Gaussian peak width.

\begin{table*}[]
\caption{Tabulation of peak location $\lambda_p$, and fractional peak area for each material's room temperature emission spectra determined using multi-peak fitting. DA = Dy:Y(acac)$_3$, DAP = Dy:Y(acac)$_3$(phen), DHD = Dy:Y(hfa)$_3$(DPEPO), D4T = Dy:Y(4-BBA)$_3$(TPPO), DAD = Dy:Y(acac)$_3$(DPEPO)}
\label{tab:peaks}
\begin{tabular}{c|cccccccccc}
\hline \\
\textbf{Transition} & \multicolumn{2}{c}{\textbf{DA}}  & \multicolumn{2}{c}{\textbf{DAP}} & \multicolumn{2}{c}{\textbf{DAD}} & \multicolumn{2}{c}{\textbf{DHD}} & \multicolumn{2}{c}{\textbf{D4T}} \\
                    & $\lambda_p$ (nm) & \textit{Frac. Area} & $\lambda_p$ (nm)  & \textit{Frac. Area} & $\lambda_p$ (nm)  & \textit{Frac. Area} & $\lambda_p$ (nm)  & \textit{Frac. Area} & $\lambda_p$ (nm) & \textit{Frac. Area} \\ \hline
${}^4I_{15/2}\rightarrow {}^6H_{15/2}$   & 456.0      & 0.004  & 456.3 & 0.005 & 457.3  & 0.009  & 456.6 & 0.009 & 457.5  & 0.002   \\
&            &                     &            &                     &            &                     &            &                     &            &                     \\
${}^4F_{9/2}\rightarrow {}^6H_{15/2}$ & 471.1 & 0.010   & 481.6  & 0.210    & 470.0      & 0.008  & 477.4  & 0.084 & 481.4  & 0.339   \\
& 474.7      & 0.014               & 483.3      & 0.050               & 474.8      & 0.024               & 482.0      & 0.027               & 490.7      & 0.110 \\
& 481.4      & 0.134               & 489.7      & 0.046               & 481.5      & 0.156               & 487.5      & 0.105               &            &                \\
 & 487.8      & 0.133               &            &                     & 487.9      & 0.133               &            &                     &            &                     \\
 &            &                     &            &                     &            &                     &            &                     &            &                     \\
${}^4F_{9/2}\rightarrow {}^6H_{11/2}$              & 569.1      & 0.022               & 573.0      & 0.390               & 569.4      & 0.096               & 575.8      & 0.221               & 576.6      & 0.305               \\
 & 577.6      & 0.565               & 579.0      & 0.067               & 576.1      & 0.042               & 576.9      & 0.036  & 575.6      & 0.228               \\
  & 583.5      & 0.063               & 582.2      & 0.216               & 578.2      & 0.518               & 574.7      & 0.488   &            &                     \\
& 566.0      & 0.034               &            &                     &            &                     &            &                     &            &                     \\
 &            &                     &            &                     &            &                     &            &                     &            &                     \\
${}^4F_{9/2}\rightarrow {}^6H_{9/2}$ & 654.9  & 0.004  & 652.3  & 0.001  & 666.4 & 0.013 & 657.4      & 0.013 & 663.9  & 0.017   \\
 & 666.8      & 0.016               & 661.2      & 0.010               &            &                     & 666.4      & 0.018               &            &     \\ \hline               
\end{tabular}
\end{table*}

From Table \ref{tab:peaks} we find that each large peak in Figure \ref{fig:spec} is composed of several component Gaussian peaks, with the component peak locations and fractional areas depending on the ligands attached to the Dy ions. For instance, by comparing the peak decompositions of Dy:Y(acac)$_3$, Dy:Y(acac)$_3$(phen), and Dy:Y(acac)$_3$(DPEPO), we find that the absence or addition of another ligand to Dy:Y(acac)$_3$ has a large influence on the peak structure. This observation is surprising as we initially anticipated that the majority of absorption and energy transfer will be due to the acac ligand. The reason for this hypothesis is that acac is nearly resonant with our pump wavelength (355 nm), while both phen and DPEPO have absorbance peaks farther from 355 nm. Given this hypothesis we would assume that the addition (or lack thereof) of another ligand (with off-resonant absorption) would have a minimal effect on the material's peak structures. To better understand the nature of energy transfer of these materials requires a detailed study of the UV excitation spectra of these materials, which is beyond the scope of this current study.

While Table \ref{tab:peaks} considers the peak decomposition of all four observed Dy transitions, for the purpose of TCT we are primarily concerned with the ${}^4F_{9/2}\rightarrow{}^6H_{15/2}$ and ${}^4I_{15/2}\rightarrow{}^6H_{15/2}$ transitions. These transitions are more easily characterized by their peak locations (instead of using multipeak decomposition), which are tabulated in Table \ref{tab:trans}. Using these peak locations we can determine the energy splitting between the ${}^4I_{15/2}$ and ${}^4F_{9/2}$ energy levels, which is also listed in Table \ref{tab:trans}. Note that the energy differences in Table \ref{tab:trans} are larger than the expected value of 0.115 eV \cite{Carnall68.01,Carnall79.01,Bunzli10.01,Bunzli10.02}. This difference is due to the emission spectra consisting of multiple overlapping peaks corresponding to transitions between Stark shifted energy sub levels. This splitting leads to the energy level difference depending on which Stark sub-levels are dominate in the emission spectra \cite{Bunzli10.01,Bi16.01}.

\begin{table*}
\caption{Peak wavelength and lifetime at room temperature for different transitions for different Dy-complexes. Note that the error in peak wavelength is 0.3 nm.}
\label{tab:trans}
\begin{tabular}{c|cccccc}
\hline 
 \textbf{Material}     & \multicolumn{2}{c}{$\mathbf{{}^4F_{9/2}\rightarrow{}^6H_{15/2}}$}    &\multicolumn{2}{c}{$\mathbf{{}^4I_{15/2}\rightarrow{}^6H_{15/2}}$}& \multicolumn{2}{c}{$\Delta E_{peaks}$} \\ 
 &    \textit{Lifetime} ($\mu$s)      & \textit{Peak Location} (nm) &      \textit{Lifetime ($\mu$s)}      & \textit{Peak Location (nm)}  &(10$^{-3}$ eV) & cm$^{-1}$\\ \hline     
Dy:Y(acac)$_3$          &  $20.24\pm0.16$    &  484.05    &  $13.75\pm0.32$    & 455.41 & 161.1  &1299 \\    
Dy:Y(acac)$_3$(phen)    &  $6.84\pm 0.30$    &  483.40    &  $4.45 \pm 0.23$   & 455.17 & 159.1  & 1283\\
Dy:Y(acac)$_3$(DPEPO)   &  $21.34\pm 0.22$   &  484.21    &  $17.13\pm 0.32$   & 455.33 & 162.4  & 1310  \\
Dy:Y(4-BBA)$_3$(TPPO)   &  $7.30 \pm 0.40$   &  481.94    &  $6.19\pm0.57$     & 457.03 & 140.2   &1131\\
Dy:Y(hfa)$_3$(DPEPO)    &  $3.39\pm0.65$     &  482.91    &  $2.76\pm0.71$     & 455.98 & 155.1  & 1251  \\ \hline
\end{tabular}
\end{table*}

In addition to determining the peak structure of the room temperature spectra, we also measure the lifetimes of the ${}^4I_{15/2}$ and ${}^4F_{9/2}$ energy levels, with Table \ref{tab:trans} listing the lifetimes for each material. From Table \ref{tab:trans} we find that the lifetimes range from 2.76 $\mu$s up to 21.34 $\mu$s.


\subsection{Thermal Performance}
\begin{figure}
\centering
\includegraphics{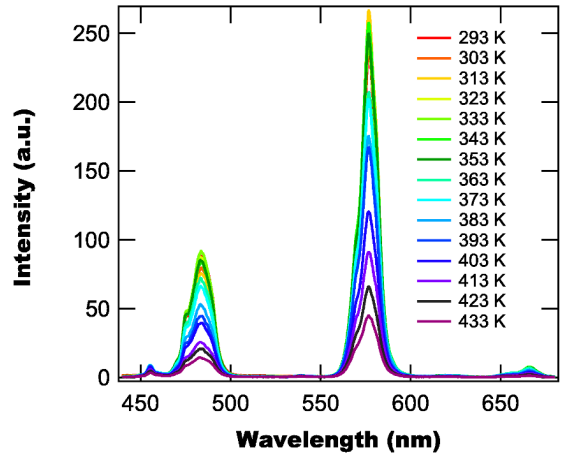}
\caption{Emission spectra of Dy:Y(acac)$_3$(DPEPO) at different temperatures. As the temperature increases the emission intensity decreases due to thermal quenching.}
\label{fig:TSpec}
\end{figure}

With the room temperature performance of the complexes determined, we next turn to consider how the complexes behave under elevated temperatures. Figure \ref{fig:TSpec} shows example emission spectra for Dy:Y(acac)$_3$(DPEPO) at different temperatures, with the emission found to decrease in intensity with increasing temperature. This decrease is a result of thermal quenching of the Dy ion's emission and is seen for all five Dy-complexes.

\begin{figure}
 \centering
 \includegraphics{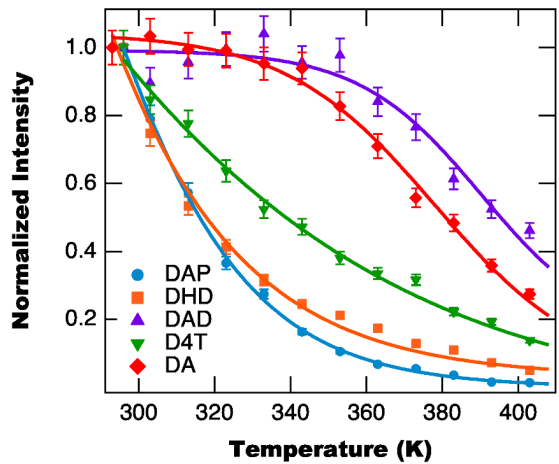}
 \caption{Intensity at 482 nm as a function of temperature, normalized such that the room temperature intensity is unity.}
 \label{fig:TQ}
\end{figure}

To determine how thermal quenching influences the emission of each complex we plot the emission intensity of the 482 nm peak as a function of temperature in Figure \ref{fig:TQ}, with the emission intensity normalized such that the room temperature intensity is unity. From Figure \ref{fig:TQ} we find that the influence of thermal quenching on each material is drastically different, with  Dy:Y(acac)$_3$(phen) found to quickly quench while the emission from Dy:Y(acac)$_3$(DPEPO) remains bright over the widest temperature range. 

To make quantitative comparisons of the effect of thermal quenching on the materials' photoluminescence intensity, we use the functional maximum temperature $T_M$ -- defined as the temperature where the $\lambda_1$ intensity is quenched to 1\% -- as our comparison parameter. We determine $T_M$ by first fitting the normalized intensities in Figure \ref{fig:TQ} to a stretched Arrhenius function:

\begin{align}
 I(T)=A\left[1-\exp\left\{-\left(\frac{T_0}{T}\right)^\beta\right\}\right], \label{eqn:strarr}
\end{align}
where $A$ is an amplitude parameter, $T_0$ is the curves characteristic temperature and $\beta$ is a stretch exponent. Note that Equation \ref{eqn:strarr} is phenomenological and should not be understood to be a fundamental relationship. Using the definition of $T_M$ (i.e. $I(T_M)=0.01$) and rearranging Equation \ref{eqn:strarr} we find that the functional maximum temperature is given by

\begin{align}
 T_M=T_0\left\{-\ln\left[1-\frac{0.01}{A}\right]\right\}^{-1/\beta}.
\end{align}
Utilizing the fit parameters calculated from Figure \ref{fig:TQ} we compute the functional maximum temperatures, which are tabulated in Table \ref{tab:temp}. Note that the $T_M$ for Dy:Y(acac)$_3$ and Dy:Y(acac)$_3$(DPEPO) are extrapolated values as they lie above the temperature range of our current heating setup (which has a maximum temperature of $\approx 473$ K). Therefore we are unable to experimentally verify the accuracy of these extrapolated values.

\begin{figure}
 \centering
 \includegraphics{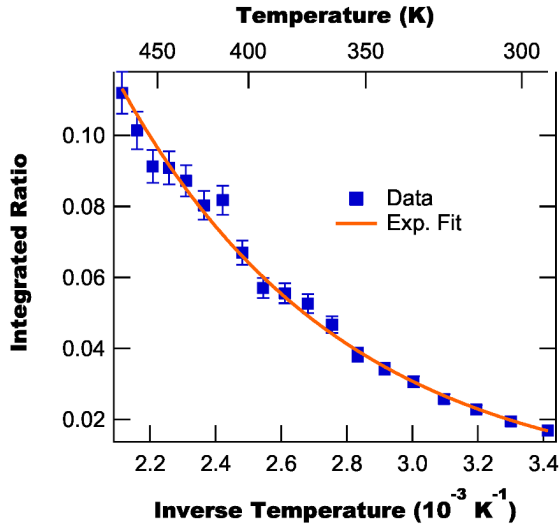}
 \caption{Integrated ratio (455/480) for Dy:Y(acac)$_3$ as a function of temperature with a fit to a simple exponential function. Table \ref{tab:temp} tabulates the exponential fit parameters for all five materials.}
 \label{fig:IR}
\end{figure}

With the effect of thermal quenching on each complex's emission quantified, we next determine the TCT calibration parameters for each complex. This is accomplished by calculating the ratio of the integrated peak intensities for the 455 nm (integration range 450-465) and 480 nm (integration range 465-504) peaks at each temperature. The uncertainty in the ratios is also computed using the uncertainty in the peak areas due to detector shot-noise and shot-to-shot laser variations. Figure \ref{fig:IR} shows an example integrated ratio as a function of inverse temperature for the Dy:Y(acac)$_3$ molecular crystals. For all five molecular crystals the integrated ratio is found to follow an exponential function as predicted by Equation \ref{eqn:fitfunc}, with Table \ref{tab:temp} tabulating the fit parameters ($A$ and $\Delta E$) for each Dy-complex. Establishing these parameters enables us to determine the molecular crystals' temperature based on their emission spectra.

\begin{table*}
\caption{Functional maximum temperatures, Integrated ratio exponential fit parameters, and maximum sensitivities for each material.}
\label{tab:temp}
\begin{tabular}{c|ccccc}
\hline 
 \textbf{Material}      & $T_M$   &     $A$        & \multicolumn{2}{c}{$\Delta E$} & $S_{MAX}$ \\
			& (K)              &                &   (10$^{-3}$ eV) &   (cm$^{-1}$)& (\% K$^{-1}$)  \\ \hline
Dy:Y(acac)$_3$          &  516   & $2.53\pm0.20$  &  $126.7\pm 2.6$  & $1022\pm21$ &1.7      \\    
Dy:Y(acac)$_3$(phen)    &  427    & $1.08\pm0.20$  & $98.4\pm6.2$    & $794\pm50$ &1.3     \\
Dy:Y(acac)$_3$(DPEPO)   & 546     & $1.83\pm0.36$  &  $117.8\pm6.6$  & $950\pm53$ &1.6 \\
Dy:Y(4-BBA)$_3$(TPPO)   & 464     & $1.19\pm0.20$  & $122.0\pm5.5$   & $984\pm44$ &1.6  \\
Dy:Y(hfa)$_3$(DPEPO)    &  433     & $0.65\pm0.27$    &  $76\pm12$    & $613\pm10$ &1.0     \\ \hline
\end{tabular}
\end{table*}

The final thermal performance parameter we consider is the maximum sensor sensitivity. To compute the maximum sensor sensitivity we substitute Equation \ref{eqn:fitfunc} into Equation \ref{eqn:sens} and find that the sensitivity as a function of temperature is, 

\begin{align}
 S(T)&=\frac{100}{r(T)}\frac{\partial r(T)}{\partial T},
\\ &=\frac{\Delta E}{kT^2}, \label{eqn:sens2}
\end{align}
which decreases with increasing temperature. Therefore, the maximum sensitivity is determined by substituting $\Delta E$ ( from Table \ref{tab:temp}) and the minimum temperature used (293 K) into Equation \ref{eqn:sens2}, with the results tabulated in Table \ref{tab:temp}. From Table \ref{tab:temp} we find that the sensitivities range from 1-1.7 \% K$^{-1}$, which are on par for other TCT phosphors \cite{Brites12.01}.

\section{Discussion}
\subsection{Ranking the complexes performance}
In the previous section we reported on the spectroscopic properties (at different temperatures) of five different Dy$^{3+}$-doped yttrium ternary complexes with the goal of determining the best complex for use as a TCT phosphor. For the purpose of comparing the performance of the complexes we consider five different criteria: (1) luminescence lifetime, (2) resistance to thermal quenching, (3) the intensity proportionality constant ($A$ in Equation \ref{eqn:fitfunc}), (4) the temperature resolution $\Delta T$, and (5) the maximum sensor sensitivity $S_M$.

For the first performance criteria we use Table \ref{tab:trans} to rank the materials in order of shortest-to-longest lifetime as: Dy:Y(hfa)$_3$(DPEPO) $<$ Dy:Y(acac)$_3$(phen) $<$ Dy:Y(4-BBA)$_3$(TPPO) $<$ Dy:Y(acac)$_3$  $<$ Dy:Y(acac)$_3$(DPEPO). Based on these results we make several important observations about the influence of the ligands on the lifetimes of the Dy excited states. First we observe that while the addition of the DPEPO ligand to the Dy:Y(acac)$_3$ system improves the material's fluorescence lifetimes, the addition of the phen ligand decreases the performance of the Dy:Y(acac)$_3$ system. This observation is consistent with previous observations on the influence of the phen ligand on Dy(acac)$_3$ \cite{Feng09.01} and on other lanthanide-doped complexes \cite{Xin03.01,Xin04.01} as well as the influence of DPEPO on other lanthanide-doped complexes \cite{Xu06.01,Biju13.01,Biju09.01,Congiu13.01}. Additionally, we find that when we substitute hfa for acac in Dy:Y(acac)$_3$(DPEPO) the fluorescence lifetime drastically decreases, despite the H-F substitution of the acac ligand. This decrease means that the stabilization of high-energy vibrations by the H-F substitution \cite{Glover07.01} is less important to Dy:Y(hfa)$_3$(DPEPO) performance than the backwards energy transfer from the Dy ion to the hfa ligand. Recall from Table \ref{tab:ligand} that hfa's lowest triplet state is only 0.18 eV greater than Dy's ${}^4F_{9/2}$ energy level. 

To determine the second performance criteria -- resistance to thermal quenching -- we use the functional maximum temperatures reported in Table \ref{tab:temp} to rank the materials (from lowest-to-highest functional maximum temperatures) as: Dy:Y(hfa)$_3$(DPEPO) $\approx$ Dy:Y(acac)$_3$(phen) $<$ Dy:Y(4-BBA)$_3$(TPPO) $<$ Dy:Y(acac)$_3$  $<$ Dy:Y(acac)$_3$(DPEPO), where Dy:Y(hfa)$_3$(DPEPO) and Dy:Y(acac)$_3$(phen) have approximately the same thermal performance. Comparing this ranking of functional maximum temperatures to the ranking we have for the room temperature lifetimes, we find that the two quantities appear to be correlated, with longer lifetimes corresponding to higher functional maximum temperatures. 

While a wider range of materials needs to be tested to confirm a relationship between the lifetime and the maximum functional temperature, this result suggests that we can roughly estimate a Dy-doped yttrium complex's resistance to thermal quenching based on the room temperature lifetime. This relationship can be intuitively argued as follows: In general, the rate of energy loss $k$, from the excited Dy energy levels is given by
\begin{align}
k=k_{rad}+k_{NR}, \label{eqn:rate}
\end{align}
where $k_{rad}$ is the rate of radiative energy loss and $k_{NR}$ is the total rate of non-radiative energy loss, which is the sum of all non-radiative channels. Equation \ref{eqn:rate} can be inverted to give the lifetime $\tau$ of the excited state to be,

\begin{align}
\tau&=\frac{1}{k}, \\ \nonumber
&=\frac{1}{k_{rad}+k_{NR}}.
\end{align}
Assuming that the radiative lifetime $\tau_{rad}=1/k_{rad}$ is similar for all complexes, the only way to change the lifetime becomes the non-radiative energy loss channels. This implies that materials with lower lifetimes have an increased number of energy loss channels. Since the non-radiative energy loss rate is known to increase with temperature, it makes intuitive sense that complexes having more loss channels will be more susceptible to thermal quenching than those with few loss channels. This is because elevating the temperature will result in a greater fraction of energy being diverted to non-radiative channels than in the case of few loss channels.

The next performance criteria is the amplitude parameter in Equation \ref{eqn:fitfunc}. From Table \ref{tab:temp} we can rank the materials as follows: Dy:Y(hfa)$_3$(DPEPO) $<$ Dy:Y(acac)$_3$(phen) $<$ Dy:Y(4-BBA)$_3$(TPPO) $<$ Dy:Y(acac)$_3$(DPEPO)  $<$ Dy:Y(acac)$_3$. From this ranking we find that both Dy:Y(acac)$_3$ and Dy:Y(acac)$_3$(DPEPO) are once again the best performing materials.

\begin{figure}
\centering
\includegraphics{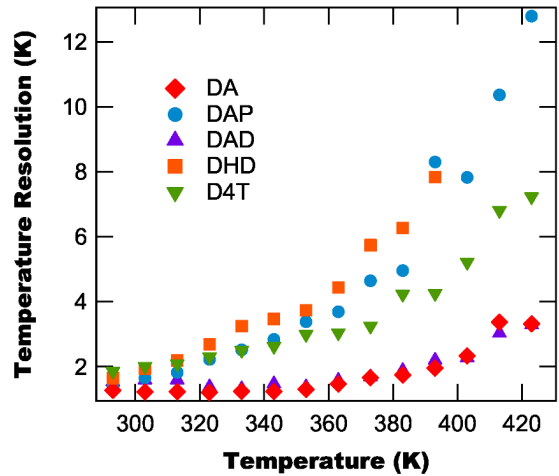}
\caption{Temperature resolution as a function of temperature for each Dy-doped yttrium complex.}
\label{fig:Tres}
\end{figure}

When determining the fourth performance criteria (temperature resolution) we find it to be temperature dependent and we therefore  plot the calculated resolution as a function of temperature in Figure \ref{fig:Tres}. From Figure \ref{fig:Tres} we find that all five sensors have similar temperature resolutions at room temperature (in the 1.5 K to 2 K range), but the resolution's behavior changes drastically as the temperature increases. Looking at the high temperature region in Figure \ref{fig:Tres} we find that we can rank the sensor's temperature resolution performance as: Dy:Y(hfa)$_3$(DPEPO) $<$ Dy:Y(acac)$_3$(phen) $<$ Dy:Y(4-BBA)$_3$(TPPO) $<$ Dy:Y(acac)$_3$(DPEPO)  $<$ Dy:Y(acac)$_3$, which is consistent with the ranking of the other metrics.

Finally, the last performance criteria for the sensors are their temperature sensitivity, which are tabulated in Table \ref{tab:temp}, with their values ranging between 1 \% K$^{-1}$ and 1.7 \% K$^{-1}$. Ranking the sensors from least-sensitive to most-sensitive we find: Dy:Y(hfa)$_3$(DPEPO) $<$ Dy:Y(acac)$_3$(phen) $<$ Dy:Y(4-BBA)$_3$(TPPO) $\approx$ Dy:Y(acac)$_3$(DPEPO)  $<$ Dy:Y(acac)$_3$, where once again Dy:Y(acac)$_3$(DPEPO) and Dy:Y(acac)$_3$ display the best performance (i.e. largest sensitivity) of the different sensors.

\subsection{Effect of ligand choice}

Based on these rankings we find that the choice of ligands not only affects the room temperature spectral performance of these complexes \cite{Bunzli10.01}, but also their usefulness as TCT phosphors, with Dy:Y(acac)$_3$ and Dy:Y(acac)$_3$(DPEPO) found to consistently outperform the other phosphors tested in this study. While a full understanding of the effect of ligand choice on the performance parameters of different complexes requires a more systematic study with a larger number of ligands, we can still make some observations based on the complexes used in this study. Specifically, by comparing Dy:Y(acac)$_3$, Dy:Y(acac)$_3$(phen) and Dy:Y(acac)$_3$(DPEPO) we can consider the influence of an auxiliary ligand (none, phen, or DPEPO) on the thermal performance.

Looking at the rankings above for Dy:Y(acac)$_3$, Dy:Y(acac)$_3$(phen) and Dy:Y(acac)$_3$(DPEPO), we find that overall Dy:Y(acac)$_3$(DPEPO) performs better than Dy:Y(acac)$_3$, which performs better than Dy:Y(acac)$_3$(phen). This implies that the DPEPO ligand improves the performance of the Dy complex, while the phen ligand decreases the performance. These results are consistent with previous studies involving these ligands where phen was observed to decrease the performance of ternary complexes \cite{Feng09.01,Xin03.01,Xin04.01}, while DPEPO improved the performance \cite{Xu06.01,Biju13.01,Biju09.01,Congiu13.01}. Phen's negative effect is attributed to its relatively low triplet level \cite{Xin03.01,Feng09.01}, while DPEPO's enhancing effect is attributed to its compact rigid structure \cite{Xu06.01}. This structure helps with energy transfer to the lanthanide ion, suppress non-radiative energy loss modes, and reduces the formation of exciplexes, which can decrease the luminescence efficiency of the complex.

Lastly, we find -- when considering Dy:Y(hfa)$_3$(DPEPO) -- that the substitution of acac with hfa drastically reduces the performance of the material.  Given this observation, we conclude that while the H-F substitution (acac to hfa) has been shown to improve luminescence performance in other complexes \cite{Glover07.01}, this improvement is mitigated in Dy:Y(hfa)$_3$(DPEPO) by backwards energy transfer from the Dy ion to the ligand. This backwards energy transfer occurs as the energy difference between the hfa's lowest triplet level and Dy's ${}^4F_{9/2}$ energy level is $< 0.23$ eV (see Table \ref{tab:ligand}).

\section{Conclusions}
We measure the temperature dependent photoluminescence of five different Dy$^{3+}$-doped yttrium complexes to determine the best material for use as a TCT phosphor. These complexes include: Dy:Y(acac)$_3$(phen), Dy:Y(hfa)$_3$(DPEPO), Dy:Y(4-BBA)$_3$(TPPO), Dy:Y(acac)$_3$, and Dy:Y(acac)$_3$(DPEPO). All five materials are found to suffer from thermal quenching at elevated temperatures, with Dy:Y(acac)$_3$(DPEPO) having the highest functional maximum temperature (546 K) and Dy:Y(acac)$_3$(phen) having the lowest functional maximum temperature (427 K).  In addition to measuring the materials' functional maximum temperature, we also determine each material's TCT calibration parameters, temperature resolution, and temperature sensitivity.

From these results we find that the  performance of the complexes depends strongly on the choice of ligands, with Dy:Y(acac)$_3$(DPEPO) having the best overall performance and Dy:Y(hfa)$_3$(DPEPO) having the worst overall performance. Also, we find that there is a correlation between the room temperature lifetime and the maximum functional temperature of the complexes, with increased lifetimes producing higher maximum functional temperatures. This result provides a guide for choosing future Dy-complexes for TCT phosphors such that the influence of thermal quenching is minimized.

To better understand how different ligand properties influence the complex's thermal performance we are currently performing a more systematic experimental study using a wider variety of ligands. This study, with systematic variations of the ligands, will help to elucidate what ligand properties produce the best material performance. It will also help elucidate the relationship between the functional maximum temperature and the room temperature lifetime. We are also planning on performing modeling of the complexes using Gaussian 09 \cite{Gaussian09}.

In addition to our results revealing new avenues of study for phosphor development, they also determine the best TCT phosphor molecular crystal (of the five tested) for use in future dynamic shock experiments, namely Dy:Y(acac)$_3$(DPEPO). Having determined our best phosphor we are now planning a series of experiments to test the phosphor's response under rapid heating and pressure, which are the two main components of dynamic shock compression. These experiments include: performing TCT measurements under rapid CO$_2$ laser heating, photoluminescence measurements under static pressure, and -- eventually -- photoluminescence measurements under dynamic shock compression.

\section*{Acknowledgements}
This work was supported by the Air Force Office of Scientific Research, Award \# FA9550-15-1-0309  to Washington State University.

\end{document}